\documentclass[conference,10pt,letterpaper]{IEEEtran}
\IEEEoverridecommandlockouts
\usepackage[T1]{fontenc}
\usepackage{graphicx}
\usepackage{amsmath,amssymb,booktabs,cite,balance}
\usepackage{tikz,pgfplots}
\usetikzlibrary{arrows.meta,positioning,patterns}
\pgfplotsset{compat=1.18}
\PassOptionsToPackage{hyphens}{url}
\usepackage{xurl}
\usepackage{eso-pic}
\usepackage[hidelinks]{hyperref}
\hypersetup{pdftitle={RetainZ: Reclaim-Time-Aware Placement for AI Checkpoints on Zoned SSDs},pdfauthor={Minxing Chu and Ruoxi Yang}}
\definecolor{ink}{HTML}{253746}
\definecolor{fifo}{HTML}{707B86}
\definecolor{coarse}{HTML}{287DA8}
\definecolor{rz}{HTML}{D87921}
\definecolor{good}{HTML}{168374}
\definecolor{dead}{HTML}{B65B35}

\makeatletter
\renewcommand{\@IEEEauthorblockconfadjspace}{-0.85em}
\renewcommand{\@IEEENORMtitlevspace}{0.6\baselineskip}
\renewcommand{\@IEEEMINtitlevspace}{0.45\baselineskip}
\def\@IEEEaftertitletext{\vspace{-5pt}}
\makeatother
\title{RetainZ: Reclaim-Time-Aware Placement\\for AI Checkpoints on Zoned SSDs%
\thanks{Code and reproducibility artifacts:
\url{https://github.com/DonaldLucy/RetainZ}.}}
\author{\IEEEauthorblockN{Minxing Chu\textsuperscript{1,2}\qquad Ruoxi Yang\textsuperscript{3}}
\IEEEauthorblockA{\begin{tabular}{@{}c@{\qquad}c@{\qquad}c@{}}
\textsuperscript{1}Harvard University & \textsuperscript{2}Brown University & \textsuperscript{3}Tufts University\\
Cambridge, MA, USA & Providence, RI, USA & Medford, MA, USA
\end{tabular}\\
minxing\_chu@\{g.harvard.edu, brown.edu\}\qquad ryang11@tufts.edu}}
\begin{document}
\maketitle
\AddToShipoutPictureFG*{%
  \AtPageLowerLeft{%
    \raisebox{24pt}[0pt][0pt]{%
      \makebox[\paperwidth][c]{%
        \parbox[b]{0.90\textwidth}{%
          \centering\fontsize{8}{9}\selectfont
          This work has been submitted to IEEE for possible publication. Copyright may be transferred without notice, after which this version may no longer be accessible.
        }%
      }%
    }%
  }%
}

\setlength{\abovedisplayskip}{4pt plus 1pt minus 1pt}
\setlength{\belowdisplayskip}{4pt plus 1pt minus 1pt}
\setlength{\abovedisplayshortskip}{0pt plus 1pt}
\setlength{\belowdisplayshortskip}{3pt plus 1pt minus 1pt}

\begin{abstract}
Solid State Drives (SSDs) are becoming the dominant medium for
performance-critical storage. Zoned Namespace (ZNS) SSDs are getting more and
more attractive because sequential writes and explicit zone resets reduce
address-mapping, over-provisioning, and internal garbage-collection costs.
However, reclaiming space requires resetting an entire zone, so deleting one
file does not free its space while other files there must be kept.
This erase-before-write constraint becomes a bottleneck for write-intensive
workloads, particularly AI checkpointing during model training, where frequent
saves of model and optimizer state compete for space with long-retained
checkpoints. We observe that retention policies inherently encode when data
can be retired, avoiding the need for a learned lifetime predictor.
We present RetainZ, a storage backend that translates these
policies into explicit reclaim epochs. Its allocator isolates long-retained data and groups objects with
nearby reclaim epochs, balancing delayed space reclamation against
unused space at zone ends.
Evaluated on FEMU with multi-tenant streams and checkpoint manifests up to Pythia-1B, RetainZ reduces space occupied by deleted data and unused zone tails by 41.7\% and mean per-input checkpoint p99 latency by 47.5\% at 84\% target load compared with the allocator of ZenFS, an established lifetime-aware ZNS backend, and lowers host write amplification from 1.165 to 1.003.
\end{abstract}
\vspace{-5pt}
\vspace{-5pt}

\suppressfloats[t]
\begin{figure}[t]
    \centering
    \includegraphics[width=\columnwidth]{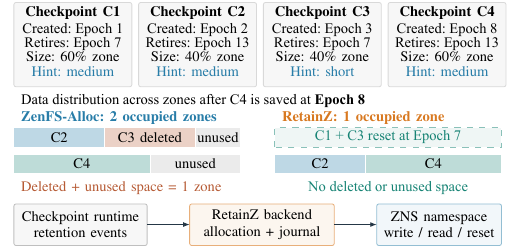}
    \caption{RetainZ retains the same live data in fewer occupied zones.}
    \label{fig:lifecycle}
    \vspace{-16pt}
\end{figure}

\section{Introduction and motivation}
The demand for high bandwidth and low latency has made Solid State Drives
(SSDs) a mainstream component of modern data centers~\cite{han2021ssd}. However,
delivering that performance through a conventional block interface incurs
costs that arise from how flash stores data. NAND flash is programmed in pages
but erased in blocks. As a result, supporting arbitrary overwrites requires address translation, spare capacity, and garbage collection that copies valid data before erasure.
These overheads limit sustained write performance and reduce the storage
capacity available to applications.

Zoned Namespace (ZNS) SSDs offer an attractive way to reduce these overheads.
The interface partitions the logical address space into zones and requires
sequential writes within each zone. Host software selects the zone for each
write and must reset an entire zone before reusing its written space.
This division of responsibility reduces the SSD controller's memory
requirements and the capacity reserved for internal management, while
reducing I/O delays caused by device-side garbage collection under
sustained write-intensive workloads~\cite{bjorling2021zns}.

However, reclaiming entire zones becomes costly when files with different
lifetimes are stored together. Deleting a short-lived file does not free
its space if another file in the same zone must be retained. To reclaim
that space, the backend must wait until the remaining files can
be deleted or copy them elsewhere before resetting the zone. As the device
fills, the shortage of free zones and the bandwidth spent on copying
can significantly delay incoming writes. This constraint is particularly relevant to AI checkpointing during model
training. Frequent saves of model and optimizer state generate large,
mostly sequential writes~\cite{aws2025}. Recent recovery checkpoints are replaced as training progresses,
while selected snapshots are retained much longer. Therefore, the backend must support rapid turnover of
large files alongside long-retained data, making timely space reclamation
essential for successive saves.
\looseness=-1

Prior work provides complementary optimizations at the training-runtime
and storage levels. At the runtime level, these optimizations reduce
training pauses and storage traffic during checkpoint creation. CheckFreq shortens these
pauses by allowing checkpoint state copying and storage writes to proceed
concurrently with training computation. It also adjusts the number of
training iterations between saves to keep checkpoint overhead within a
target budget~\cite{mohan2021checkfreq}. FastPersist writes checkpoint
data to multiple SSDs in parallel and allows independent training
computations to continue during these writes~\cite{fastpersist}.
Reducing saved data also lowers this
overhead. Check-N-Run uses differential checkpointing to save modified
model state and quantization to represent saved values with fewer
bits~\cite{eisenman2022checknrun}. 
\looseness=-1

At the storage level, prior work improves file placement and zone
reclamation to reduce the cost of recovering usable space
~\cite{kim2026lrza,zarc,han2021znsplus}. Placing files expected to be
deleted at similar times in the same zone leaves less live data to
copy when that zone is reclaimed. ZenFS, a RocksDB storage backend
for ZNS SSDs, groups files using coarse lifetime hints that distinguish
short-lived from longer-lived data~\cite{zenfs}. Prophet estimates
when individual database files will be deleted and groups files with
similar predicted deletion times~\cite{prophet}. 

However, these methods do not explicitly use the runtime retention
rules that determine when a checkpoint can be deleted to guide
placement.
Our key insight is that, for AI checkpointing, the training runtime
can provide these deletion conditions before checkpoint data are
written. The conditions follow directly from the runtime's retention
policy, which specifies the events that make each generation eligible
for deletion. Exposing them to the storage backend allows it to
choose which generations share zones according to their retirement
conditions, grouping data that can be reclaimed together.
\looseness=-1

Training frameworks such as Megatron and TorchTitan support policies
that retain a limited number of recent checkpoint generations
~\cite{megatron,torchtitan}.
Under a policy that keeps the three most recent completed generations,
generation 17 becomes eligible for deletion once generation 20 has
been durably saved. The runtime can specify this condition when
generation 17 is created and pass it to the storage backend with
the write request. Using these retirement conditions effectively requires balancing
timely reclamation with efficient use of zone capacity. Assigning
each generation to separate zones makes its space independently
reclaimable, but can leave unused space at zone ends and consume
the device's limited active-zone slots. Sharing zones improves
packing, yet combining generations with distant retirement events
delays space reuse. 

We present RetainZ, a zoned storage backend that connects runtime
retirement conditions to zone placement and reclamation.
Figure~\ref{fig:lifecycle} compares RetainZ with the ZenFS allocator
(ZenFS-Alloc), illustrating how grouping checkpoints
that retire together allows entire zones to be reclaimed without
relocating live data. RetainZ implements this principle through
an explicit runtime--storage contract and an online allocator
that balances reclamation delay against unused space at zone ends.
\looseness=-1

\vspace{-2pt}
\section{Retention-Guided Placement}
\label{sec:retention_design_v2}
\vspace{-3pt}

Runtime retention rules determine when individual checkpoint
generations can be deleted, while the storage backend reclaims
their space by resetting entire zones. RetainZ is a host-side software backend that uses these rules
to plan which generations share a zone
(Figure~\ref{fig:retention_placement}(a)). This requires considering
both how long deleted data remain in the zone and how much
capacity is left unused.

\subsection{From retention rules to reclaim hints}
\vspace{-2pt}

When creating a checkpoint, the runtime supplies its birth epoch $a$
on the checkpoint workload's logical timeline, shard sizes, and a
reclaim hint $d$. We define this hint as the epoch associated with
the runtime event that makes the generation eligible for deletion
under its retention policy. Under a latest-$K$ policy, which retains the $K$ most recent completed
generations, generation $g$ becomes eligible for deletion when
generation $g+K$ has been durably committed. Checkpoints selected for long-term
retention beyond this window are referred to as milestones. RetainZ
represents them as anchors with $d=\infty$, indicating that no
retirement epoch is specified. For a declared checkpoint cadence $c$, latest-$K$ retention gives
$d=a+cK$ on the shared logical timeline, allowing cross-stream
comparison of policy-derived reclaim hints.
\looseness=-1

The backend uses these hints to guide placement and tracks durable
commit and deletion events to determine when space can be reclaimed.
A generation's stored extents remain live until the runtime explicitly
deletes the generation and the deletion is durably recorded.
When no live extents remain in a zone, the backend can reset that
zone and reuse its capacity.

\vspace{-2pt}
\subsection{Accounting for reclamation delay and unused space}
\vspace{-2pt}

With reclaim hints available, the allocator needs a way to compare
the capacity costs of different placements. An occupied zone can hold little
live data because much of its contents have been deleted or because it was
never filled. RetainZ accounts for both sources of overhead and how long
they persist. Let $U_z$ and $V_z$ denote written and live bytes in a zone
of capacity $C$. Deleted data occupy $U_z-V_z$ bytes, and the unwritten
tail occupies $C-U_z$ bytes. Over an observation interval of $T$ checkpoint
epochs, their combined cost is
\begin{equation}
 \mathcal{B}=\int_0^T\!\sum_{z\;\mathrm{occupied}}
 \big[(U_z-V_z)+(C-U_z)\big]\,dt.
 \label{eq:reserved_overhead_v2}
\end{equation}
The first term accumulates dead-byte time (DBT), while the second accounts
for unused tails over the same interval. Dividing
$\mathcal{B}$ by $CT$ gives the mean capacity reserved beyond live
data, expressed in units of zone capacity. Subsequent appends can fill a writable tail and reduce this overhead.
FINISH prevents further appends to the zone, so any remaining tail
stays unavailable until RESET.

To guide an individual write, RetainZ compares the incoming reclaim
hint with those of the data already written to a candidate zone.
For finite hints, let $D$ denote the maximum reclaim hint among
the extents written to that zone. Appending $n$ bytes
with hint $d$ changes this maximum to $D'=\max(D,d)$.
The marginal DBT cost computed from these hints is
\begin{equation}
 \Delta\mathrm{DBT}=n(D'-d)+U_z(D'-D).
 \label{eq:delta_dbt_v2}
\end{equation}
The first term charges the incoming bytes for waiting beyond their own
retirement epoch. The second charges the zone's existing written bytes for extending
the maximum reclaim hint from $D$ to $D'$. This includes bytes that are
invalid, because postponing RESET also prolongs their occupation
of the zone. RetainZ combines this marginal cost with a tail-capacity
term to approximate~\eqref{eq:reserved_overhead_v2}, so that avoiding
retirement mismatch does not simply leave more zones partially filled.

\vspace{-2pt}
\subsection{Selecting zones for incoming writes}
\vspace{-2pt}

RetainZ keeps anchors in separate zones so that checkpoints with
finite reclaim hints can be reclaimed independently of them.
Anchor placement favors zones with less unused capacity after
the write. For finite hints, the allocator first limits the spread
of retirement epochs within a zone and then compares the capacity
costs of the resulting candidates.

For an incoming checkpoint with birth epoch $a$ and finite reclaim
hint $d$, the retention horizon is $h=d-a$. Each zone holding
finite hints maintains their minimum $m$, maximum $D$, and the
shortest retention horizon $H$ among its written extents.
RetainZ sets a base span constraint using the shorter retention
horizon:
\begin{equation}
 \max(D,d)-\min(m,d)
 \leq \max\{1,\lfloor\min(H,h)/2\rfloor\}.
 \label{eq:span_v2}
\end{equation}
For a generation of padded size $S<C/2$, the allocator uses
the larger of the limit in \eqref{eq:span_v2} and
$(\lfloor C/S\rfloor-1)\max\{1,\lfloor\min(H,h)/4\rfloor\}$.
This admits a wider range of retirement epochs when several
small generations can fit in one zone, improving opportunities
to fill unused tails. When generations are deleted at their
declared epochs, the resulting span bounds the interval between
the first and last deletion, limiting how long deleted data
wait for the zone to become reclaimable.
\looseness=-1

\begin{figure}[t]
    \centering
    \includegraphics[width=\columnwidth]
        {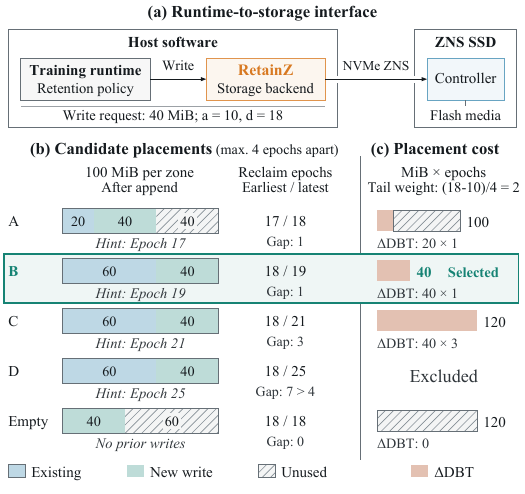}
    \vspace{-22pt}
    \caption{RetainZ architecture and zone selection.}
    \label{fig:retention_placement}
    \vspace{-10pt}
\end{figure}

Among compatible zones and available empty zones, the allocator
uses the cost model to decide where to write next. For a shard
with $s$ aligned bytes remaining, a candidate can receive
$n=\min\{s,C-U_z\}$ bytes and is assigned the score
\begin{equation}
 J(z)=\Delta\mathrm{DBT}+\alpha(C-U_z-n),
 \qquad \alpha=\max\{1,h/4\}.
 \label{eq:score_v2}
\end{equation}
The first term accounts for the additional waiting caused by
combining retirement epochs, as defined in
\eqref{eq:delta_dbt_v2}. The second expresses the remaining tail
capacity in byte-epochs, making it comparable to
$\Delta\mathrm{DBT}$. Its weight $\alpha$ grows with $h$ because
capacity reserved by checkpoints with longer retention horizons
can remain unused for longer. An empty zone has $U_z=0$ and
$\Delta\mathrm{DBT}=0$, so its score reflects the capacity left
unused after accepting the write. The allocator selects the
lowest score, breaking ties in favor of the zone with fewer
host RESETs. If a shard spans multiple zones, the allocator
repeats this selection for its remaining bytes. Figure~\ref{fig:retention_placement}(b)--(c) shows an example
with $h=8$ and $H\ge8$ for all occupied candidates, yielding
a four-epoch span limit.
\looseness=-1

Device resource limits create an additional reason to use
existing tails. Opening an empty zone at the active-zone limit
requires finishing an active zone, which can strand its remaining
tail. When no candidate is available under the normal placement
rule, RetainZ can relax the span constraint to fill existing tails
while continuing to keep anchors separate from data with finite
hints. The same cost function ranks these fallback candidates.
This trades a wider spread of retirement epochs for access to
capacity already available in occupied zones.
Section~\ref{sec:refinements} further examines how checkpoint size
and zone resource limits affect these placement decisions.

\section{Online Backend and Evaluation Method}
\label{sec:method}

We implement RetainZ in C++17 to evaluate how its placement
decisions affect space reuse and checkpoint performance.
The backend combines direct I/O, a durable journal, and cached
per-zone hint summaries. Our primary baseline is ZenFS-Alloc,
a port of the upstream ZenFS data-file allocator~\cite{zenfs}.
Equal-epoch best-fit receives the same reclaim hints as RetainZ
but groups matching retirement epochs during normal placement;
ablations remove either cost term or the span constraint.
All policies replay identical workloads and share the I/O path,
journal, copying GC, and zone-resource budgets. The pressure sweep reserves one empty zone for GC.
Additional controls use a common admission rule that fills
compatible tails and borrows the reserved zone only when
the remaining padded checkpoint data fit in it.

Experiments run on FEMU~\cite{li2018femu} under QEMU CPU emulation.
The pressure sweep uses a 256-MiB namespace with 4-MiB zones,
leaving one open and six active zones for data after shared
resource reservations. Four synthetic streams combine
latest-$K$ retention with staggered saves and nominal checkpoint
sizes of 0.74--2.27 zone capacities. Each input requests
250 saves at a minimum 100-ms start-to-start interval.
Resident anchors set the target peak live occupancy after
scheduled deletions to 70--90\%. We complement these controlled workloads with model-derived
experiments. Capacity tests replay original Pythia-1B checkpoint
shard lengths using verifiable pattern bytes on 64--256-GiB
namespaces~\cite{pythia}. These runs use lossless compressed
backing, while latency and real-payload recovery tests use
DRAM backing. Real-payload tests write SmolLM2-135M checkpoints
captured during CPU training with PyTorch Distributed
Checkpoint (DCP), using retention events from the upstream
Megatron and TorchTitan routines~\cite{smollm2,dcp,megatron,torchtitan}.
After backend recovery, DCP reloads model, AdamW, and RNG
state. We compare the restored state and next training update
with a reference run.

We report reserved overhead $\mathcal{B}/(CT)$, host write
amplification (WA; host-written bytes per logical checkpoint byte),
and backend p99 latency (request handling through durable commit,
excluding serialization). Synthetic metrics exclude prefill and
the first 20 epochs. Pressure results show means and min--max
across three seeds, averaging per-input p99.

\begin{figure}[t]
    \centering
    \includegraphics[width=0.85\columnwidth]
        {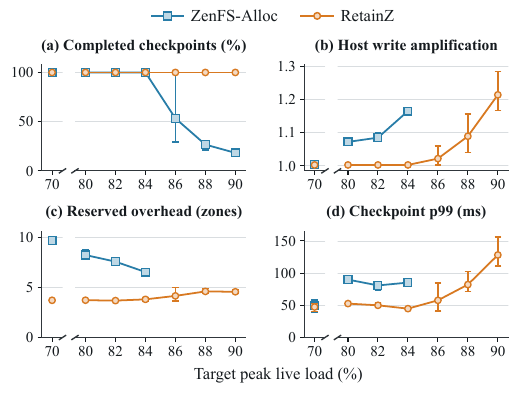}
    \vspace{-14pt}
    \caption{FEMU results with a fixed GC reserve.}
    \vspace{-12pt}
    \label{fig:pressure}
\end{figure}

\section{Measured Behavior}
\label{sec:refinements}
\textbf{Placement Effectiveness.}
RetainZ sustains checkpoint writes with lower capacity overhead
and less copying. Figure~\ref{fig:pressure} shows results
under increasing capacity pressure with a fixed GC reserve.
At 84\% load, RetainZ reduces reserved overhead from 6.53 to
3.80 zones relative to ZenFS-Alloc, host WA from 1.165 to
1.003, and checkpoint p99 latency from 85.3 to 44.8 ms.
RetainZ avoids live-data copying at this load, whereas
ZenFS-Alloc requires foreground migration on some commits.
At 90\% load, RetainZ completes 250 saves per input,
while ZenFS-Alloc exhausts capacity after 40--49 saves.

\begin{figure}[t]
    \centering
    \includegraphics[width=\columnwidth]
        {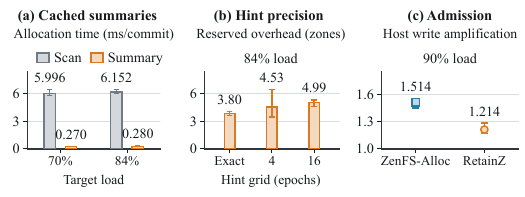}
    \vspace{-24pt}
    \caption{Mechanism controls for RetainZ.}
    \vspace{-16pt}
    \label{fig:operating}
\end{figure}

With common admission, both allocators complete all requested
saves on the same 90\%-load inputs.
Figure~\ref{fig:operating}(c) shows that RetainZ still reduces
mean host WA from 1.514 to 1.214. Its placement therefore
reduces copying even when both policies have sufficient
headroom to complete each replacement.

\textbf{Ablation and Sensitivity.}
Table~\ref{tab:placement_controls} examines where these gains
come from under common admission. Invalid and Tail separate
reserved overhead into deleted data and unused zone tails.
The placement ablations average five matched inputs; the
remaining controls average three new inputs. With identical reclaim hints, RetainZ reduces reserved
overhead by 34.1\% relative to equal-epoch best-fit at 84\%
load, showing the benefit of choosing zone sharing through
both reclamation and packing costs. Removing the tail term
leaves more unused capacity, while removing the DBT term
leaves more invalid data. Isolating every generation
exhausts capacity in this workload.
\looseness=-1

\begin{table}[t]
\centering
\caption{Placement quality and sensitivity.}
\label{tab:placement_controls}
\vspace{-6pt}
\begingroup
\footnotesize
\setlength{\tabcolsep}{2.25pt}
\renewcommand{\arraystretch}{1.03}
\begin{tabular}{@{}lrrrr@{}}
\toprule
Policy or setting & Invalid & Tail & Host WA & $p_{99}$ \\
 & (zones) & (zones) & & (ms) \\
\midrule
\multicolumn{5}{@{}l}{\textit{Placement and ablations, 84\% load}} \\
ZenFS-Alloc         & 3.66 & 3.11 & 1.153 & 79.8 \\
Equal-epoch best-fit & 2.96 & 2.82 & 1.045 & 63.4 \\
DBT only            & 1.61 & 3.61 & 1.014 & 48.4 \\
Tail only, same span  & 2.35 & 2.10 & 1.007 & 46.9 \\
No span constraint  & 1.69 & 1.88 & 1.003 & 43.9 \\
\textbf{RetainZ}     & 1.11 & 2.70 & 1.003 & 43.6 \\
\midrule
\multicolumn{5}{@{}l}{\textit{Span control: RetainZ / no span}} \\
90\% load
 & 3.54/3.95 & 1.00/0.64 & 1.189/1.251 & 119.5/127.7 \\
Small checkpoints
 & 0.81/1.83 & 2.04/1.64 & 1.008/1.008 & 16.8/15.8 \\
Delayed deletion
 & 1.41/1.89 & 2.99/2.36 & 1.003/1.003 & 45.7/39.8 \\
Extended retention
 & 2.17/3.05 & 3.30/2.18 & 1.003/1.004 & 42.2/44.4 \\
\midrule
\multicolumn{5}{@{}l}{\textit{Scale/resources: RetainZ / ZenFS-Alloc}} \\
4 MiB, 6 active
 & 1.06/3.68 & 2.79/3.13 & 1.003/1.123 & 40.3/72.5 \\
4 MiB, 14 active
 & 1.11/3.99 & 2.75/4.85 & 1.003/1.644 & 40.7/158.1 \\
4 MiB, 30 active
 & 1.11/6.34 & 2.75/3.66 & 1.003/2.080 & 37.9/202.7 \\
64 MiB, 6 active
 & 1.06/3.68 & 2.79/3.13 & 1.000/1.120 & 517.2/836.5 \\
\bottomrule
\vspace{-22pt}
\end{tabular}
\endgroup
\end{table}

The span constraint trades some packing efficiency for less
invalid data trapped in occupied zones. Removing it lowers
total overhead at 84\% load and in the delayed-deletion and
extended-retention tests at 80\% load. At 90\% load, retaining
the constraint reduces mean host WA by 4.9\%, as tighter
capacity makes copying more costly. Size adaptation helps
recover packing efficiency for small checkpoints: across
five matched inputs at 70\% load, it reduces reserved
overhead by 11.5\% relative to the same allocator without
size adaptation.

Figure~\ref{fig:operating}(b) examines hint precision in the
fixed-reserve sweep. At 84\% load, rounding finite hints
upward to 4- and 16-epoch grids increases reserved overhead
from 3.80 to 4.53 and 4.99 zones, respectively, while
checkpoint sizes and actual deletion events remain unchanged.
More precise hints allow the allocator to distinguish
retirement events that coarse hints group together. The final block of Table~\ref{tab:placement_controls} tests
zone size and resource budgets. Scaling zones from 4 to
64 MiB, with namespace and checkpoint sizes increased
proportionally, preserves RetainZ's advantage: reserved
overhead and p99 latency are 43.4\% and 38.2\% lower than
ZenFS-Alloc's. With one data-open slot, increasing the
data-active budget from 6 to 14 and 30 slots leaves
RetainZ's overhead near 3.86 zones. ZenFS-Alloc selects
fresh zones more often as additional slots become available,
increasing fragmentation and copying.

\textbf{Allocation Overhead and Recovery Validation.}
Cached zone summaries make these placement decisions
inexpensive. Figure~\ref{fig:operating}(a) shows: at
70\% load, replacing full-history scans with cached summaries
reduces mean allocation cost from 5.996 to 0.270 ms per
commit, including summary maintenance, while preserving
placement decisions and write traffic. Measured allocator metadata totals 430.2~KiB for both
RetainZ and ZenFS-Alloc. Both allocators complete the 64-GiB capacity trials using
original Pythia-1B shard lengths under common admission,
with readback verifying every retained byte.
SmolLM2-135M payload tests additionally verify recovery
through DCP under both Megatron and TorchTitan retention
policies. DCP restores model, AdamW, and RNG state, and the
next training update matches the reference loss, parameters,
and optimizer state. Recovery also succeeds in process-crash
tests at commit, relocation, and FINISH boundaries.
\looseness=-1

\section{Conclusion}
RetainZ turns checkpoint retention policies into placement
information for zoned storage. By carrying retirement conditions
from the runtime to the backend, it coordinates zone sharing
with reclamation while accounting for unused capacity.
Therefore, application semantics can guide storage decisions
before data are written, allowing placement and space reuse
to be designed together throughout each checkpoint's lifecycle.

\balance
\bibliographystyle{IEEEtran}
\bibliography{references}

@inproceedings{bjorling2021zns,
  author={Matias Bj{\o}rling and Abutalib Aghayev and Hans Holmberg and Aravind Ramesh and {Le Moal}, Damien and Gregory R. Ganger and George Amvrosiadis},
  title={{ZNS}: Avoiding the Block Interface Tax for Flash-based {SSDs}},
  booktitle={USENIX Annual Technical Conference (ATC)},year={2021},pages={689--703},
  url={https://www.usenix.org/conference/atc21/presentation/bjorling}}

@misc{zenfs,
  author={{Western Digital}},title={{RocksDB} with {ZenFS}},year={2026},
  url={https://zonedstorage.io/docs/applications/zenfs},note={Accessed September 26, 2026}}

@inproceedings{kim2026lrza,
  author={Jinyoung Kim and Doeun Kim and Junmo Seong and Hyuck Han and Sooyong Kang},
  title={Lifetime-Aware Zone Allocation for {ZNS SSDs}},
  booktitle={IEEE International Conference on Fog and Edge Computing (ICFEC)},
  year={2026},pages={89--90},doi={10.1109/ICFEC69006.2026.00019}}

@misc{megatron,
  author={{NVIDIA}},title={{Megatron-LM}: Checkpointing implementation},year={2026},
  url={https://github.com/NVIDIA/Megatron-LM/blob/d113016bb2850c4b7d804f24ba72c055d7d4f861/megatron/training/checkpointing.py},
  note={Revision d113016, accessed September 26, 2026}}

@misc{torchtitan,
  author={{PyTorch}},title={{TorchTitan}: Checkpointer retention policy},year={2026},
  url={https://github.com/pytorch/torchtitan/blob/35ceeb5901e6ffe6a0a3ecd4c4e6e9e80cbc853e/torchtitan/components/checkpointer/base.py},
  note={Revision 35ceeb5, accessed September 26, 2026}}

@misc{aws2025,
  author={Venkata Sistla and Yanjie Niu and Hossam Basudan and Siva Devabakthini},
  title={Architecting scalable checkpoint storage for large-scale {ML} training on {AWS}},
  year={2025},month=jun,
  url={https://aws.amazon.com/blogs/storage/architecting-scalable-checkpoint-storage-for-large-scale-ml-training-on-aws/}}

@inproceedings{li2018femu,
  author={Huaicheng Li and Mingzhe Hao and Michael Hao Tong and Swaminathan Sundararaman and Matias Bj{\o}rling and Haryadi S. Gunawi},
  title={The {CASE} of {FEMU}: Cheap, Accurate, Scalable and Extensible Flash Emulator},
  booktitle={USENIX Conference on File and Storage Technologies (FAST)},
  year={2018},pages={83--90},url={https://www.usenix.org/conference/fast18/presentation/li}}

@misc{smollm2,
  author={Ben Allal, Loubna and others},
  title={{SmolLM2}: When Smol Goes Big---Data-Centric Training of a Small Language Model},
  year={2025},eprint={2502.02737},archivePrefix={arXiv},url={https://arxiv.org/abs/2502.02737}}

@misc{dcp,
  author={{PyTorch}},title={Distributed Checkpoint},year={2025},
  url={https://docs.pytorch.org/docs/2.8/distributed.checkpoint.html},
  note={Experiments use PyTorch 2.8.0}}

@inproceedings{zarc,
  author={Farhan S. Chowdhury and Shadman Saqib Eusuf and Subhadeep Sarkar and Tarikul Islam Papon},
  title={{ZARC}: Re-inventing Zone Allocation for {LSMs} on {ZNS SSD}},
  booktitle={International Workshop on Data Management on New Hardware (DaMoN)},
  year={2026},doi={10.1145/3789237.3809132},
  url={https://www.cs.umb.edu/~tpapon/pdfs/zarc_damon_cr.pdf}}

@inproceedings{prophet,
  author={Gaoji Liu and Chongzhuo Yang and Qiaolin Yu and Chang Guo and Wen Xia and Zhichao Cao},
  title={{Prophet}: Optimizing {LSM}-based Key-Value Store on {ZNS SSDs} with File Lifetime Prediction and Compaction Compensation},
  booktitle={38th International Conference on Massive Storage Systems and Technology (MSST)},
  year={2024},
  url={https://www.msstconference.org/MSST-history/2024/Papers/msst24-3.1.pdf}}

@misc{fastpersist,
  author={Guanhua Wang and Olatunji Ruwase and Bing Xie and Yuxiong He},
  title={{FastPersist}: Accelerating Model Checkpointing in Deep Learning},
  year={2024},eprint={2406.13768},archivePrefix={arXiv},
  doi={10.48550/arXiv.2406.13768},url={https://arxiv.org/abs/2406.13768}}

@misc{pythia,
  author={Stella Biderman and Hailey Schoelkopf and Quentin Anthony and Herbie Bradley and Kyle O'Brien and Eric Hallahan and Mohammad Aflah Khan and Shivanshu Purohit and USVSN Sai Prashanth and Edward Raff and Aviya Skowron and Lintang Sutawika and Oskar van der Wal},
  title={{Pythia}: A Suite for Analyzing Large Language Models Across Training and Scaling},
  year={2023},eprint={2304.01373},archivePrefix={arXiv},
  doi={10.48550/arXiv.2304.01373},url={https://arxiv.org/abs/2304.01373}}

@inproceedings{han2021znsplus,
 author={Kyuhwa Han and Hyunho Gwak and Dongkun Shin and Jooyoung Hwang},
 title={{ZNS+}: Advanced Zoned Namespace Interface for Supporting In-Storage Zone Compaction},
 booktitle={USENIX Symposium on Operating Systems Design and Implementation (OSDI)},year={2021},pages={147--162},
 url={https://www.usenix.org/conference/osdi21/presentation/han}}

@inproceedings{mohan2021checkfreq,
 author={Jayashree Mohan and Amar Phanishayee and Vijay Chidambaram},
 title={{CheckFreq}: Frequent, Fine-Grained {DNN} Checkpointing},
 booktitle={USENIX Conference on File and Storage Technologies (FAST)},year={2021},pages={203--216},
 url={https://www.usenix.org/conference/fast21/presentation/mohan}}

@inproceedings{eisenman2022checknrun,
 author={Assaf Eisenman and Kiran Kumar Matam and Steven Ingram and Dheevatsa Mudigere and Raghuraman Krishnamoorthi and Krishnakumar Nair and Misha Smelyanskiy and Murali Annavaram},
 title={{Check-N-Run}: A Checkpointing System for Training Deep Learning Recommendation Models},
 booktitle={USENIX Symposium on Networked Systems Design and Implementation (NSDI)},year={2022},pages={929--943},
 url={https://www.usenix.org/conference/nsdi22/presentation/eisenman}}

@inproceedings{han2021ssd,
 author={Shujie Han and Patrick P. C. Lee and Fan Xu and Yi Liu and Cheng He and Jiongzhou Liu},
 title={An In-Depth Study of Correlated Failures in Production {SSD}-Based Data Centers},
 booktitle={USENIX Conference on File and Storage Technologies (FAST)},
 year={2021},pages={417--429},
 url={https://www.usenix.org/conference/fast21/presentation/han}}

\end{document}